\documentclass[11pt,a4paper]{article}
\usepackage{amsfonts}
\usepackage{amsbsy}
\usepackage{amssymb}
\usepackage{amsmath}
\usepackage{mathrsfs}
\usepackage{latexsym}
\usepackage{color}
\usepackage{revsymb}

\newcommand{\be}{\begin{equation}}
\newcommand{\ee}{\end{equation}}
\newcommand{\ba}{\begin{eqnarray}}
\newcommand{\ea}{\end{eqnarray}}
\newcommand{\Gammabol}{\Gamma{}}
\newcommand{\Rbol}{R{}}
\newcommand{\nabol}{\nabla{}}
\def\onehalf{{\textstyle{\frac{1}{2}}}}

\def\mt{{\mbox{\tiny{(1)}}}}
\def\my{{\mbox{\tiny{(2)}}}}

\begin{document}

\renewcommand{\thefootnote}{\fnsymbol{footnote}}
\noindent
{\Large \bf Nonlinear Gravitational Waves: Their Form and Effects}
\vskip 0.7cm
\noindent
{\bf R. Aldrovandi,$^{1}$ 
J. G. Pereira,$^{1}$ 
R. da Rocha$^{2}$ 
and K. H. Vu$^{1}$} 
\vskip 0.3cm \noindent
$^1${\it Instituto de F\'{\i}sica Te\'orica},
{\it Universidade Estadual Paulista} \\
{\it Rua Dr.\ Bento Teobaldo Ferraz 271},
{\it 01140-070 S\~ao Paulo, Brazil}

\vskip 0.3cm \noindent
$^2${\it Centro de Matem\'atica, Computa\c c\~ao e Cogni\c c\~ao}\\
{\it Universidade Federal do ABC, 09210-170 Santo Andr\'e, Brazil}
\vskip 0.8cm
\begin{quote}
{\bf Abstract}~{\footnotesize A gravitational wave must be nonlinear to be able to transport its own source, that is, energy and momentum. A physical gravitational wave, therefore, cannot be represented by a solution to a linear wave equation. Relying on this property, the second-order solution describing such physical waves is obtained. The effects they produce on free particles are found to consist of nonlinear oscillations along the direction of propagation.}
\end{quote}


\section{Introduction}
\setcounter{footnote}{0}
\renewcommand{\thefootnote}{\arabic{footnote}}

For a long period after the advent of general relativity, the question of the existence or not of gravitational waves was a very controversial issue. In the seventies, the discovery of a binary pulsar system whose orbital period changes according to the predicted wave emission put an end to the controversy~\cite{HT}. In fact, that discovery provided an indirect but compelling experimental evidence for the existence of gravitational waves~\cite{maggiore,bina}. That evidence, however, did not provide any clue on their form and effects. As a matter of fact, although widely considered a finished topic~\cite{giants}, it actually remains plagued by many obscure points~\cite{cs1}. For example, although there seems to be an agreement that the transport of energy-momentum by gravitational waves is essentially a nonlinear phenomenon, instead of going to the second order, one usually assumes a ``mixed'' procedure, which consists basically in assuming that gravitational waves carry energy (are nonlinear, or at least second order), but at the same time, because this energy is very small, one also assumes that its evolution can approximately be described by a linear (first order) equation~\cite{weinberg}. When one speaks of ``linear gravitational waves", therefore, one means nonlinear gravitational waves whose dynamics is assumed to be approximately described by a linearized equation. This means that, in addition to the sequential levels of accuracy implied by the perturbative analysis, there is also another approximation, implied by the ``mixed'' approach, according to which all first-order equations describing a gravitational wave are to be interpreted as only nearly correct~\cite{mtw}.

Such assumption, however, is an unjustified surmise: the issue is not a matter of approximation, but a conceptual question.\footnote{There are other arguments against this assumption. See, for example, Ref.~\cite{salerno}.} A gravitational wave either does or does not carry energy: if it carries, no matter how small it is, it cannot satisfy a linear equation. It is, therefore, conceptually unsatisfactory to assume that a gravitational wave satisfying a linear equation is able to transport energy and momentum. If applied to a Yang--Mills propagating field~\cite{itzu}, the approximation described above would correspond to assume that, for a field with small-enough amplitude, its evolution can be accurately described by a linear equation. Of course, this is plainly wrong: a Yang-Mills propagating field must necessarily be nonlinear to carry its own (color) source, otherwise it is not a Yang-Mills field. Analogously, a gravitational wave must necessarily be nonlinear to transport its own source---that is, energy and momentum.

Taking into account these premises, a critical review of the gravitational wave theory has been published recently~\cite{waves}. In that paper, it was discussed why the standard approach to the gravitational wave theory is not satisfactory. Here, instead of using the mixed approach, we proceed to the second order and obtain the corresponding nonlinear gravitational wave. It is important to remark that this re-interpretation of the gravitational wave concept has no implications for the usual expressions of the power emitted by a mechanical source. In particular, the (nonlinear) quadrupole radiation formula gives a correct account of the energy emitted by a binary pulsar, for example. The only we claim is that the energy and momentum are not transported away by linear, but by nonlinear waves. The basic purpose of the present paper is to make an analysis of these waves, as well as of their effects on test particles.

\section{Linear Approximation}

\subsection{Linear Wave Equation}

The study of gravitational waves involves basically the weak field approximation 
of Einstein equation
\be
\Rbol_{\mu \nu} -  \onehalf  \, g_{\mu \nu} \, \Rbol =
\frac{8 \pi G}{c^4} \, \Theta_{\mu \nu},
\ee
where $\Theta_{\mu \nu}$ is the source energy-momentum tensor. That is arrived at  
by expanding the metric tensor according to
\be
\sqrt{-g}g^{\mu \nu} = \eta^{\mu \nu} + \varepsilon \, h_\mt^{\mu \nu} +
\varepsilon^2 \, h_\my^{\mu \nu} + \dots \,,
\label{mexpansion}
\ee
where $\varepsilon$ is a small parameter introduced to label the successive 
orders in this perturbation scheme. When the metric tensor is expanded according to 
(\ref{mexpansion}), we are automatically assuming that there is a background 
Minkowskian structure in spacetime, with metric $\eta_{\mu \nu}$. Accordingly, 
the gravitational waves are interpreted as perturbations
\be
h^{\mu \nu} = \varepsilon \, h_\mt^{\mu \nu} +
\varepsilon^2 \, h_\my^{\mu \nu} + \dots 
\ee
propagating on that fixed Minkowskian background. This interpretation is 
consistent with general relativity, as well as with the point of view of field 
theory, according to which a field always propagates on a background 
spacetime~\cite{living}.

Assuming expansion (\ref{mexpansion}), the first order Ricci tensor is
\be
\Rbol_{\mt \mu \nu} = \partial_\lambda \Gammabol_\mt^\lambda{}_{\mu \nu} -
\partial_\nu \Gammabol_\mt^\lambda{}_{\mu \lambda} \, .
\ee
Using the first order Christoffel connection
\be
\Gammabol_\mt^\lambda{}_{\mu \nu} = \onehalf \left(
- \partial_\mu h_\mt^\lambda{}_\nu - \partial_\nu h_\mt^\lambda{}_\mu +
\partial^\lambda h_{\mt \mu \nu}
+ \onehalf \delta^\lambda_\nu \, \partial_\mu h_\mt 
+ \onehalf \delta^\lambda_\mu \, \partial_\nu h_\mt
- \onehalf \eta_{\mu\nu} \, \partial^\lambda h_\mt
 \right) , 
\label{Christo}
\ee
the Ricci tensor and the scalar curvature become, respectively,
\be
\Rbol_{\mt \mu \nu} = \onehalf \left(\Box h_{\mt \mu \nu} - 
\onehalf\eta_{\mu\nu}\Box h_\mt -
\partial_\lambda \partial_\mu h_\mt^\lambda{}_\nu -
\partial_\nu \partial^\lambda h_{\mt \mu \lambda} \right)
\ee
and
\be
\Rbol_\mt = - \onehalf\Box h_\mt - \partial_\lambda \partial_\rho h_\mt^{\lambda 
\rho} \, ,
\ee
where $\Box = \, \eta^{\rho \lambda} \, \partial_\rho \partial_\lambda$ is the 
flat spacetime d'Alembertian, and $h_\mt = h_\mt^{\lambda}{}_{\lambda}$. In 
consequence, the first order sourceless gravitational field equation
becomes
\begin{equation}
\Box h_{\mt\mu\nu} - \partial_\lambda\partial_\mu h_\mt^\lambda{}_\nu 
- \partial_\nu\partial^\lambda h_{\mt\mu\lambda} + 
\eta_{\mu\nu}\partial_\lambda\partial_\rho h_\mt^{\lambda\rho} = 0.
\label{eqs4}
\end{equation}

Now, as is well known, wave equation (\ref{eqs4}) is invariant under (infinitesimal) general 
spacetime coordinate transformations. Analogously to the electromagnetic wave 
equation, which is invariant under gauge transformations, the ambiguity of the 
gravitational wave equation can be removed by choosing a particular class of 
coordinate systems~---~or gauge,  as it is usually called. The most convenient 
choice is the class of harmonic coordinate systems, which at first order is 
fixed by
\be
\partial_\mu h_\mt^{\mu\nu} = 0.
\label{hcc}
\ee
In this case, the field equation (\ref{eqs4}) reduces to the relativistic wave 
equation
\be
\Box \, h_\mt^\rho{}_\nu = 0.
\label{we1}
\ee

\subsection{Linear Waves}
\label{TTcs}

A monochromatic plane-wave solution to equation (\ref{we1}) has the form
\be
h_{{\mbox{\tiny{(1)}}}\mu \nu} = A_{{\mbox{\tiny{(1)}}}\mu \nu}  \exp[i k_{\rho} 
x^\rho],
\label{pw}
\ee
where $A_{{\mbox{\tiny{(1)}}}\mu \nu}=A_{{\mbox{\tiny{(1)}}}\nu \mu}$ is the 
polarization tensor, and the wave vector $k^\rho$ satisfies
\be\label{12}
k_{\rho} \, k_{}^\rho = 0.
\ee
The harmonic coordinate condition (\ref{hcc}), on the other hand, implies
\be\label{13}
k_\mu \, h_\mt^\mu{}_\nu = 0.
\ee

In analogy with the Lorentz gauge in electromagnetism, it is possible to 
further specialize the harmonic class of coordinates to a particular coordinate 
system. Once this is done, the coordinate system becomes completely specified, 
and the components $A_{\mt \mu \nu}$ turn out to represent only physical degrees 
of freedom. A quite convenient choice is the so called {\it transverse--traceless} coordinate system (or gauge), in which~\cite{schutz}
\be
h_{\mbox{\tiny{(1)}}}^\rho{}_\rho = 0 \quad \mbox{and} \quad
h_{\mbox{\tiny{(1)}}}^\mu{}_\nu \, U_{\mbox{\tiny{(0)}}}^\nu = 0,
\label{TTA}
\ee
with $U_{\mbox{\tiny{(0)}}}^\nu$ an arbitrary, constant four-velocity.

Now, although the coordinate system $\{x^\mu\}$ has already been completely 
specified (the transverse-traceless coordinate system), we still have the 
freedom to choose different local Lorentz frames $e^a$. Since the metric $g_{\mu 
\nu} = \eta_{ab} \, e^a{}_\mu e^b{}_\nu$ is invariant under changes of frames, 
the metric perturbation will also be invariant. In particular, it is always 
possible to choose a specific frame, called {\it proper 
frame}, in which $U_{\mbox{\tiny{(0)}}}^\nu = \delta_0{}^\nu$. In this frame, as 
can be seen from the second of the Eqs.~(\ref{TTA}),
\be
h_{\mbox{\tiny{(1)}}}^\mu{}_0 = 0
\ee
for all $\mu$. Linear waves satisfying these conditions are usually assumed to 
represent a plane gravitational wave in the transverse-traceless gauge, 
propagating in the vacuum with the speed of light. Its physical significance, 
however, can only be determined by analyzing the energy and momentum it 
transports.

\subsection{Energy and Momentum Transported by Linear Waves}

The energy-momentum tensor of any matter (or source) field $\psi$ is proportional to the functional derivative of the corresponding Lagrangian with respect to  the spacetime metric. Since such a derivative does not change the order of the Lagrangian in the matter field $\psi$, both the Lagrangian and the energy-momentum tensor will be of the same order in the field 
variable $\psi$. For example, both Maxwell's Lagrangian and its  corresponding energy-momentum tensor are quadratic in the electromagnetic field. Now, it is a well known fact that the gravitational field is itself a source of gravitation. This means that the gravitational energy-momentum current should 
appear explicitly in the gravitational field equation. Accordingly, the wave equation (\ref{we1}) should read
\be
\Box \, h_{\mt \mu \nu} = \frac{16 \pi G}{c^4} \, t_{\mt \mu \nu}.
\label{we1bis}
\ee
At the {\em linear ap\-prox\-imation}, therefore, the gravitational energy--momentum density $t_{\mt \mu \nu}$ is restricted to be linear. However, since the energy--momentum density is at least quadratic in the field variable, $t_{\mt \mu \nu}$ vanishes in the linear approximation, leading to the wave equation~(\ref{we1}).

The above property is a crucial difference between linear gravity and electromagnetism, and is often a source of confusion. Even though the electromagnetic waves are linear, they do transport energy and momentum. There is no any inconsistency in this result because neither energy nor momentum are sources of electromagnetic field, and consequently the energy-momentum tensor does not appear explicitly in the electromagnetic field equation. In other words, even though the electromagnetic field equations are linear, the energy-momentum tensor is not restricted to be linear. The linearity of the electromagnetic wave equation, however, restricts the electromagnetic self--current to be linear, and consequently to vanish. This means that the electromagnetic wave is unable to transport its own source, that is, electric charge. A linear gravitational wave is similarly unable to transport its own source, that is, energy and momentum. Only a nonlinear wave will be able to do it. This a subtle, but fundamental difference between electromagnetic and gravitational waves.

The consistency of this result can be verified by analyzing the generation of 
linear waves. In the presence of a source, the first order field equation reads
\be
\Box \, h_{\mt \mu \nu} =  \frac{16 \pi G}{c^4} \; \Theta_{\mt \mu \nu},
\label{fo}
\ee
with $\Theta_{\mt \mu \nu}$ the first order source energy-momentum tensor. As a consequence of the coordinate condition (\ref{hcc}), it is easy to see that
\be
\partial^\mu {\Theta}_{{\mbox{\tiny{(1)}}}\mu \nu} = 0.
\label{focl}
\ee
Instead of the usual covariant derivative, 
$\Theta_{{\mbox{\tiny{(1)}}} \mu \nu}$ is conserved with an ordinary derivative 
at the first order. Since this is a true conservation law, in the sense that it 
leads to a time conserved {\it charge}, we can conclude that in the linear 
approximation a mechanical system cannot lose energy in the form of 
gravitational waves.\footnote{This is consistent with the fact that linear 
gravitational waves do not transport energy nor momentum. The existence of a linear solution is a mere consequence of the use of a perturbative scheme, but alone it does not represent the physical wave.} As discussed in section~1, this problem is usually circumvented by assuming the mixed approach, according to which this equation is to be interpreted as nearly true.

\subsection{Generation of Linear Waves}

Let us consider the first order field equation (\ref{fo}). A solution is the retarded potential
\be
h_{\mt \mu \nu} =  \frac{4 G}{c^4} \int \frac{d^3 x'}{|\vec{x} - \vec{x}{\,}'|} \,
{\Theta}_{{\mbox{\tiny{(1)}}}\mu \nu}(t', \vec{x}{\,}'),
\label{genera1}
\ee
with the source considered in the retarded time
\be
t' = t - \frac{|\vec{x} - \vec{x}{\,}'|}{c}.
\ee
At large distances from the source we can expand
\be
|\vec{x} - \vec{x}{\,}'| \simeq r - \vec{x}{\,}' \cdot \hat{n} + \dots ,
\ee
where $r = |\vec{x}|$ is the distance from the source, and $\hat{n}$ is a unit vector in the direction of $\vec{x}$. The leading order term of $h_{\mt \mu \nu}$ is obtained by replacing $|\vec{x} - \vec{x}{\,}'|$ in the denominator of Eq.~(\ref{genera1}) with $r$,
\be
h_{\mt \mu \nu} =  \frac{4 G}{r c^4} \int {d^3 x'} \,
{\Theta}_{{\mbox{\tiny{(1)}}}\mu \nu}(t', \vec{x}{\,}'),
\label{gene1}
\ee
where now
\be
t' = t - \frac{r}{c} + \frac{\vec{x}{\,}' \cdot \hat{n}}{c}.
\ee
The Fourier transform of ${\Theta}_{{\mbox{\tiny{(1)}}}\mu \nu}$ is
\be
{\Theta}_{{\mbox{\tiny{(1)}}}\mu \nu}(t', \vec{x}{\,}') = 
\int \frac{d^4k}{(2 \pi)^4} \; \tilde{\Theta}_{{\mbox{\tiny{(1)}}}\mu \nu}(\omega, \vec{k}) \,
e^{-i \omega t' + i \vec{k} \cdot \vec{x}'}.
\ee
Substituting in Eq.~(\ref{gene1}) and performing the integrations in $d^3x'$ and $d^3k$, we obtain~\cite{maggiore} 
\be
h_{\mt \mu \nu} =  \frac{4 G}{r c^4} \int_{0}^{\infty} \frac{dw}{2 \pi} \,
\tilde{\Theta}_{{\mbox{\tiny{(1)}}}\mu \nu}(\omega, \omega{\,} \hat{n}/c )
e^{-i \omega (t - r/c)}.
\label{gene2}
\ee
We see from this expression that, if the source oscillates with a single frequency $\omega$, the plane wave $h_{\mt \mu \nu}$ will necessarily propagate with the same frequency. However, we know from the quadrupole radiation formula that, if the source oscillates with frequency $\omega$, the gravitational radiation should come out with frequency $2 \omega$.\footnote{See, for example, Ref.~\cite{maggiore}, page 105.} The reason for this factor of 2 is that both the generation and the effects of gravitational waves on free particles are essentially tidal effects, which we know to occur twice during a complete cycle. This is a clear indication that $h_{\mt \mu \nu}$ alone cannot represent the physical gravitational wave.

\section{Second Order Approximation}

\subsection{Second--Order Wave Equation}

At the second order of the iterated perturbation scheme, the harmonic coordinate 
condition reads~\cite{living}
\be
\partial_\mu h_{\mbox{\tiny{(2)}}}^{\mu}{}_{\nu} = 0.
\label{harmo2}
\ee
In these coordinates, the second order gravitational field equation can be 
written in the form
\be
\Box \, h_{\my}^{\rho}{}_{\nu} = \frac{16 \pi G}{c^4} \left( t_{\my}^{\rho}{}_{\nu} 
+ \Theta_{\my}^{\rho}{}_{\nu} \right),
\label{sofe1}
\ee
where $t_{\my}^{\rho}{}_{\nu} \equiv t_{\my}^{\rho}{}_{\nu}(h_\mt, h_\mt)$ 
represents all terms coming from the left-hand side of Einstein equation, in 
addition to the d'Alembertian term. It can be interpreted as the second order energy-momentum 
pseudotensor of the gravitational field~\cite{weinberg}.

Far away from the sources, the second order gravitational waves are governed by 
the sourceless version of the wave equation (\ref{sofe1}),
\be
\Box \, h_{\my}^{\mu \nu} = \frac{16 \pi G}{c^4} \, t_{\my}^{\mu \nu}
\equiv N^{\mu \nu}(h_\mt, h_\mt),
\label{sofeless}
\ee
where, already considering the traceless gauge condition $h_\mt = 0$, 
\begin{eqnarray}
N^{\mu \nu} (h_\mt, h_\mt) =-h^{\rho\sigma}_\mt \partial_\rho \partial_\sigma h^{\mu\nu}_\mt+\frac{1}{2} \partial^\mu h_{\mt\rho\sigma} \partial^\nu h^{\rho\sigma}_\mt+\partial_\sigma h^{\mu \rho}_\mt
(\partial^\sigma h^\nu_{\mt\rho}+ \partial_\rho h^{\nu \sigma}_\mt) ~~~~~~~~~~  \nonumber \\
- \partial^{\mu}h_{\mt\rho\sigma} \partial^\rho h_\mt^{\nu\sigma}
- \partial^{\nu}h_{\mt\rho\sigma} \partial^\rho h_\mt^{\mu \sigma} +
\frac{\eta^{\mu\nu}}{2} \bigg(\partial_\rho h_{_\mt\sigma\tau}
\partial^\sigma h^{\rho \tau}_{\mt}-\frac{1}{2} \partial_\tau h_{\mt\rho\sigma} 
\partial^\tau h_\mt^{\rho\sigma} \bigg).
\label{iteracao1}
\end{eqnarray}
Using for $h^{\mu\nu}_\mt$ the plane wave solution (\ref{pw}), the wave equation becomes
\begin{eqnarray}\label{15}
\Box\,  h_\my^{\mu\nu} = \bigg[
k_\rho k_\sigma A^{\rho\sigma}_\mt A^{\mu\nu}_\mt
- \frac{1}{2} k^\mu k^\nu A_{\mt\rho\sigma} A^{\rho\sigma}_\mt
- k_\sigma  k^\sigma A^{\mu \rho}_\mt A^\nu_{\mt\rho} ~~~~~~~~ \nonumber \\
-\, k_\sigma k_\rho A^{\mu \rho}_\mt  A^{\nu \sigma}_\mt
+ k^\nu k^\rho A_{\mt\rho\sigma}  A^{\mu \sigma}_\mt  
+ k^\mu k^\rho A_{\mt\rho\sigma}  A_\mt^{\nu\sigma}~~~~~~~~ \nonumber \\
- \frac{\eta^{\mu\nu}}{2} \Big(k_\rho k^\sigma A_{\mt\sigma\tau}
A^{\rho \tau}_\mt-\frac{1}{2} k_\tau k^\tau A_{\mt\rho\sigma}  A^{\rho\sigma}_\mt \Big) \bigg]
\exp[ i 2 k_\rho x^\rho].
\label{iteracao2}
\end{eqnarray}
Use of the constraints (\ref{12}) and (\ref{13})  reduces it to
\begin{equation}
\Box\,  h_\my^{\mu\nu} = - \frac{\Phi_\my}{2} \, k^\mu k^\nu
\exp[ i {\,}2{\,} k_\rho x^\rho], 
\label{ffee}
\end{equation}
with
\be
\Phi_\my = A_{\mt\rho\sigma}  A^{\rho\sigma}_\mt.
\label{fi2}
\ee
It is worth mentioning that the second-order wave equation is quadratic in the first-order solution $h^{\mu}_{\mt \nu}$. The factor ``2'' in the exponential of the right-hand side is a reminder of this  nonlinear, quadratic dependence.

\subsection{Second--Order Nonlinear Waves}

A general solution to the wave equation (\ref{ffee}) is given by a solution to the homogeneous equation plus a particular solution to the non--homogeneous equation. A monochromatic traveling--wave solution can then be written in the form 
\be
h_\my^{\mu\nu} = \left(A_\my^{\mu\nu} + i B_\my^{\mu\nu} \right)
\exp[ i {\,}2 k_\rho x^\rho],
\label{PhysGraWave}
\ee
where
\be
A_\my^{\mu\nu} = - \, \frac{\Phi_\my}{16} \, \eta^{\mu \nu}
\ee
and
\be
B_\my^{\mu\nu} = \frac{\Phi_\my}{8} \, \frac{K_\theta x^\theta}{K_\sigma k^\sigma} \, k^\mu k^\nu,
\label{B2}
\ee
with $K_\alpha$ an arbitrary wave number four--vector. As a direct inspection shows, this solution satisfies the harmonic coordinate condition (\ref{harmo2}). The physical gravitational wave is represented by the real part of the solution, that is,
\be
h_\my^{\mu\nu} = A_\my^{\mu\nu} \, \cos[2 k_\rho x^\rho] -
B_\my^{\mu\nu} \, \sin[2 k_\rho x^\rho].
\label{PhysGraWaveBis}
\ee
Observe that the amplitude $B_\my^{\mu\nu}$ depends explicitly on the wave number~---~or equivalently, on the frequency of the wave. This is a typical property of nonlinear waves.

The amplitude of the first part of the solution satisfies
\be
A_\my^{\mu}{}_{\mu} \equiv A_\my = -\, \frac{\Phi_\my}{4} \quad \mbox{and} \quad
k_\mu A_\my^{\mu\nu} = \frac{1}{4} \, k^\nu A_\my.
\ee
We consider now a laboratory frame~---~with a Cartesian coordinate system~---~from which the wave will be observed. In this case, only the diagonal components of $A_\my^{\mu\nu}$ are non--vanishing and obey
\be
A^{xx} = A^{yy} = A^{zz} = -\, A^{tt}.
\ee
More specifically,
\begin{gather}
\left( A_\my^{\mu \nu} \right) = - \frac{\Phi_\my}{16} \left(
\begin{matrix}
 1 & 0 & 0 & 0 \\
 0 &-1 & 0 & 0 \\
 0 & 0 & -1 & 0 \\
 0 & 0 & 0 & -1
\end{matrix}
\right).
\end{gather}
The second part, on the other hand, satisfies
\be
B_\my^{\mu}{}_{\mu} \equiv B_\my = 0 \quad \mbox{and} \quad k_\mu B_\my^{\mu\nu} = 0.
\ee
If we consider, for example, a wave traveling in the $z$ direction of the Cartesian system, for which
\be
k^\rho = \left({\omega}/{c}, 0, 0, {\omega}/{c} \right),
\label{kzdir}
\ee
the coefficient $B_\my^{\mu \nu}$ will be of the form
\begin{gather}
\left( B_\my^{\mu \nu} \right) = \frac{\Phi_\my \, K_\theta x^\theta \, \omega^2}{8 K_\alpha k^\alpha c^2} \left(
\begin{matrix}
 1 & 0 & 0 & 1 \\
 0 & 0 & 0 & 0 \\
 0 & 0 & 0 & 0 \\
 1 & 0 & 0 & 1
\end{matrix}
\right).
\end{gather}
Considering both parts of the solution we see that the second order wave is neither transverse nor traceless.

As already seen, if $r$ denotes the distance from the source, the amplitude of the first order solution scales according to $A_\mt^{\mu\nu} \sim 1/r$. As an immediate consequence, $\Phi_\my \sim 1/r^2$. This means that the amplitude of the first part of the solution (\ref{PhysGraWaveBis}) falls off as
\be
A_\my^{\mu\nu} \sim 1/r^2.
\ee
Due to an additional linear dependence on the distance, the amplitude of the second part falls off as
\be
B_\my^{\mu\nu} \sim 1/r.
\ee
At large distances from the source, therefore, the dominant solution will be of the form
\be
h_\my^{\mu\nu} \simeq B_\my^{\mu\nu} \, \sin[2 k_\rho x^\rho].
\label{dom}
\ee
Usually, second-order effects are supposed to fall off as $1/r^2$, and for this reason they are assumed to be neglectful at large distances from the source \cite{schutz}. However, as shown above, the second-order gravitational wave $h_\my^{\mu\nu}$ falls off as $1/r$, and consequently the arguments used to neglect them are not valid in this case. Observe also that, if the source oscillates with a single frequency $\omega$, the field $h_\my^{\mu\nu}$ will propagate, as appropriate for a quadrupole radiation, with a frequency $2 \omega$ \cite{maggiore}. This factor of 2 is a direct consequence of the nonlinear nature of the gravitational wave (see the comment just below Eq.~(\ref{fi2})), and provides one more evidence that $h_\my^{\mu\nu}$~---~and not $h_\mt^{\mu\nu}$~---~represents the physical (quadrupole) gravitational wave.

\subsection{Generation of Nonlinear Waves}

As can be seen from Eqs.~(\ref{harmo2}) and (\ref{sofe1}), the second order 
total energy-momentum tensor is conserved:
\be
\partial_\mu \left[t_{\my}^{\mu}{}_{\nu} + \Theta_{\my}^{\mu}{}_{\nu} \right] = 0.
\ee
The source energy-momentum tensor, on the other hand, as determined by the 
second order Bianchi identity, is conserved only in the covariant sense:
\be
\nabla_\mu \Theta_{\my}^{\mu}{}_{\nu} \equiv
\partial_\mu \Theta_{\my}^{\mu}{}_{\nu} +
\Gammabol_\mt^\mu{}_{\rho \mu} \, \Theta_{\mt}^{\rho}{}_{\nu} -
\Gammabol_\mt^\rho{}_{\nu \mu} \, \Theta_{\mt}^{\mu}{}_{\rho} = 0.
\ee
At the second order, therefore, the source energy-momentum tensor is not truly conserved~---~it does not lead to a conserved {\it charge}. As a matter of fact, the above covariant conservation law is not a true conservation law, but simply an identity governing the exchange of energy and momentum between gravitation and matter~\cite{kopo}. As a consequence, in contrast to what happens at the first order, at the second order a mechanical system can lose energy in the form of gravitational waves.

It is important to remark once more that the usual expressions of the power emitted by a mechanical source, and in particular the quadrupole radiation formula, give a correct account of the energy emitted by a mechanical system. The reason is that nonlinear methods have always been used in the study of wave generation by such systems. Furthermore, the quadratic energy-momentum pseudotensor $t_{\my}^{\rho}{}_{\nu}$ is the complex traditionally used to calculate the energy and momentum transported by gravitational waves. What we claim here is that, instead of being transported by the linear waves $h_{\mt}^{\mu \nu}$, this energy is actually transported by the second-order gravitational wave $h_{\my}^{\mu \nu}$. Notice from Eq.~(\ref{sofe1}) that $t_{\my}^{\mu \nu}$ appears as source of the second-order gravitational field $h_{\my}^{\mu \nu}$. It represents, therefore, the energy and momentum transported by the second-order gravitational waves.

\section{Effects on Free Particles}

\subsection{The Geodesic Deviation Equation}

Let us consider, as usual, two nearby particles separated by the four-vector 
$\xi^\alpha$. This vector obeys the geodesic deviation equation
\be
\nabol_U \, \nabol_U \xi^\alpha = \Rbol^\alpha{}_{\mu \nu \beta} \,
U^\mu \, U^\nu \, \xi^\beta,
\label{gde0}
\ee
where $U^\mu = dx^\mu/ds$, with $ds = g_{\mu \nu} \, dx^\mu dx^\nu$, is the 
four-velocity of the particles. Now, each order of the gravitational field 
expansion 
\be
\Rbol^\alpha{}_{\mu \nu \beta} = \varepsilon \, 
\Rbol_{\mbox{\tiny{(1)}}}^\alpha{}_{\mu \nu \beta} +
\varepsilon^2 \, \Rbol_{\mbox{\tiny{(2)}}}^\alpha{}_{\mu \nu \beta} + \dots \, ,
\ee
which follows naturally from (\ref{mexpansion}), will give rise to a different 
contribution to $\xi^\alpha$. For consistence reasons, therefore, this vector 
must also be expanded:
\be
\xi^\alpha = \xi_{\mbox{\tiny{(0)}}}^\alpha + \varepsilon \, 
\xi_{\mbox{\tiny{(1)}}}^\alpha +
\varepsilon^2 \, \xi_{\mbox{\tiny{(2)}}}^\alpha + \dots \; .
\ee
In this expansion, $\xi_{\mbox{\tiny{(0)}}}^\alpha$ represents the initial, that 
is, undisturbed separation between the particles. As the four--velocity $U^\mu$ depends on the gravitational field, it should also be expanded. However, since the gravitational wave is interpreted as a perturbation of the flat Minkowski spacetime, the movement produced on free particles will also be considered in Minkowski spacetime. This means that we can write $U^\mu = U_{\mbox{\tiny{(0)}}}^\mu = 
dx^\mu/ds_{\mbox{\tiny{(0)}}}$, where
\be
ds_{\mbox{\tiny{(0)}}}^2 = \eta_{\mu \nu} \, dx^\mu dx^\nu
\label{ds0}
\ee
is the flat spacetime quadratic interval. Of course, the four-velocity $U_{\mbox{\tiny{(0)}}}^\mu$ depends on the choice of the initial condition~---~or equivalently, on the 
choice of the local Lorentz frame from which the phenomenon will be observed and 
measured. Using then the freedom to choose this frame (see section \ref{TTcs}), 
we can choose a frame fixed at one of the particles~---~called {\it proper 
frame}. In that frame, the proper time $s_{\mbox{\tiny{(0)}}}$ 
coincides with the coordinate $x^0$~\cite{mtw}, and the particle four--velocity assumes the form
\be
U_{\mbox{\tiny{(0)}}}^\mu \equiv \delta^\mu{}_0 = (1, 0, 0, 0).
\label{U0}
\ee

\subsection{First--Order Effects}

Considering that $\xi_{\mbox{\tiny{(0)}}}^\alpha$ represents simply the undisturbed separation between the particles, at the lowest order the geodesic deviation equation is
\be
\frac{d^2 \xi_{\mbox{\tiny{(1)}}}^\alpha}{ds_{\mbox{\tiny{(0)}}}^2} +
U^\rho_{\mbox{\tiny{(0)}}} \, \partial_\rho 
\left(\Gammabol_{{\mbox{\tiny{(1)}}}}^\alpha{}_{\beta \gamma} \, 
U^\gamma_{\mbox{\tiny{(0)}}} \right) \xi_{\mbox{\tiny{(0)}}}^\beta  =
\Rbol_{\mbox{\tiny{(1)}}}^\alpha{}_{\mu \nu \beta} \, U^\mu_{\mbox{\tiny{(0)}}} 
\, U^\nu_{\mbox{\tiny{(0)}}} \; \xi_{\mbox{\tiny{(0)}}}^\beta.
\label{gde01}
\ee
Substituting $U^\mu_{\mbox{\tiny{(0)}}}$ as given by Eq.~(\ref{U0}), we get
\be
\frac{d^2 \xi_{\mbox{\tiny{(1)}}}^\alpha}{ds_{\mbox{\tiny{(0)}}}^2} + \partial_0 
\Gammabol_{{\mbox{\tiny{(1)}}}}^\alpha{}_{\beta 0} \; 
\xi_{\mbox{\tiny{(0)}}}^\beta  = \Rbol_{\mbox{\tiny{(1)}}}^\alpha{}_{00\beta} \; 
\xi_{\mbox{\tiny{(0)}}}^\beta.
\label{gde2}
\ee
Using then the first order Riemann tensor
\be
\Rbol_{\mbox{\tiny{(1)}}}^\alpha{}_{\mu \nu \beta} =
\partial_\nu \Gammabol_{\mt}^\alpha{}_{\mu \beta} -
\partial_\beta \Gammabol_{{\mbox{\tiny{(1)}}}}^\alpha{}_{\mu \nu},
\label{R1}
\ee
it reduces to
\be
\frac{d^2 \xi_{\mbox{\tiny{(1)}}}^\alpha}{ds_{\mbox{\tiny{(0)}}}^2} + \partial_0 
\Gammabol_{{\mbox{\tiny{(1)}}}}^\alpha{}_{\beta 0} \; 
\xi_{\mbox{\tiny{(0)}}}^\beta  = \left( \partial_0 
\Gammabol_{{\mbox{\tiny{(1)}}}}^\alpha{}_{\beta 0} -
\partial_\beta \Gammabol_{{\mbox{\tiny{(1)}}}}^\alpha{}_{0 0} \right) 
\xi_{\mbox{\tiny{(0)}}}^\beta.
\label{gde2bis}
\ee
Canceling $\partial_0 \Gammabol_{{\mbox{\tiny{(1)}}}}^\alpha{}_{\beta 0} \; 
\xi_{\mbox{\tiny{(0)}}}^\beta$ on both sides, we get
\be
\frac{d^2 \xi_{\mbox{\tiny{(1)}}}^\alpha}{ds_{\mbox{\tiny{(0)}}}^2} = - \, 
\partial_\beta \Gammabol_{{\mbox{\tiny{(1)}}}}^\alpha{}_{0 0} \; 
\xi_{\mbox{\tiny{(0)}}}^\beta,
\label{gde3}
\ee
where
\be
\Gammabol_{\mt}^\alpha{}_{0 0} = \onehalf \, \eta^{\alpha \rho} \left( 2 \; 
\partial_0 h_{\mt\rho 0} -
\partial_\rho h_{\mt 00} \right).
\ee
Specializing now to the transverse-traceless coordinate system, where the 
components $h_{\mt \rho 0}$ vanish identically, we obtain
\be
\frac{d^2 \xi_{\mt}^\alpha}{ds_{\mbox{\tiny{(0)}}}^2} = 0.
\label{gde4A}
\ee
Without loss of generality, we can take the solution to be $\xi_{\mbox{\tiny{(1)}}}^\alpha$ = 
constant. In the linear approximation, therefore, in consonance with the fact that linear gravitational waves do not transport energy nor momentum, particles are not affected by linear gravitational waves.\footnote{For a detailed discussion of this point, see Ref.~\cite{waves}.}

\subsection{Second-Order Effects}

Up to second order, and already using the first order results, the geodesic deviation equation (\ref{gde0}) reads
\be
\frac{d^2 \xi_{\mbox{\tiny{(2)}}}^\alpha}{ds_{\mbox{\tiny{(0)}}}^2} + \Gammabol_{{\mbox{\tiny{(1)}}}}^\alpha{}_{\gamma 0} \; \Gammabol_{{\mbox{\tiny{(1)}}}}^\gamma{}_{\beta 0} \; \xi_{\mbox{\tiny{(0)}}}^\beta + \partial_0 \Gammabol_{{\mbox{\tiny{(2)}}}}^\alpha{}_{\beta 0} \; \xi_{\mbox{\tiny{(0)}}}^\beta =
\Rbol_{\mbox{\tiny{(2)}}}^\alpha{}_{0 0 \beta} \; \xi_{\mbox{\tiny{(0)}}}^\beta.
\label{gde2A}
\ee
Substituting the curvature tensor
\be
R^{\alpha}_{\my 00 \beta} = \partial_0 \Gamma^\alpha_{\my 0 \beta} -
\partial_\beta \Gamma^\alpha_{\my 0 0} +
\Gamma^\alpha_{\mt 0 \gamma} \Gamma^\gamma_{\mt 0 \beta} -
\Gamma^\alpha_{\mt \beta \gamma} \Gamma^\gamma_{\mt 0 0},
\ee
and considering that in transverse--traceless coordinates $\Gamma^\gamma_{\mt 0 0} = 0$, we obtain
\be
\frac{d^2 \xi_{\mbox{\tiny{(2)}}}^\alpha}{ds_{\mbox{\tiny{(0)}}}^2} = - \,
\partial_\beta \Gamma^\alpha_{\my 0 0} \, \xi_{\mbox{\tiny{(0)}}}^\beta.
\label{gde3A}
\ee
Now, in transverse--traceless coordinates, the second order Christoffel connection is
\begin{equation}
\Gamma^\alpha_{\my 0 0} = \partial_0 h^\alpha_{\my 0} - \onehalf
\partial^\alpha h_{\my 00}.
\label{Christo2}
\end{equation}
The geodesic deviation equation reduces then to
\be
\frac{d^2 \xi_{\mbox{\tiny{(2)}}}^\alpha}{ds_{\mbox{\tiny{(0)}}}^2} = \left(
\frac{1}{2} \partial_\beta \partial^\alpha h_{\my 00} -
\partial_\beta \partial_0 h^\alpha_{\my 0} \right) \xi_{\mbox{\tiny{(0)}}}^\beta.
\label{gde4}
\ee

For definiteness, we consider a wave traveling in the $z$ direction, in which case $k^\rho$ is given by Eq.~(\ref{kzdir}). Let us then suppose two particles separated initially in the $x$ direction by a distance $\xi_{\mbox{\tiny{(0)}}}^x$, that is,
\be
\xi_{\mbox{\tiny{(0)}}}^\beta = (0, \xi_{\mbox{\tiny{(0)}}}^x, 0, 0).
\ee
Considering that, in the proper frame $s_{\mbox{\tiny{(0)}}} = ct$, it is an easy task to verify that in this case the resulting equations of motion are
\be
\frac{\partial^2 \xi_{\mbox{\tiny{(2)}}}^x}{\partial t^2} =
\frac{\partial^2 \xi_{\mbox{\tiny{(2)}}}^y}{\partial t^2} =
\frac{\partial^2 \xi_{\mbox{\tiny{(2)}}}^z}{\partial t^2} = 0.
\label{gde5}
\ee
The same result is obtained for two particles separated initially in the $y$ direction. We consider now two particles separated initially in the $z$ direction by a distance $\xi_{\mbox{\tiny{(0)}}}^z$, that is,
\be
\xi_{\mbox{\tiny{(0)}}}^\beta = (0, 0, 0, \xi_{\mbox{\tiny{(0)}}}^z).
\ee
In this case, the geodesic deviation equation (\ref{gde4}) yields
\be
\frac{\partial^2 \xi_{\mbox{\tiny{(2)}}}^x}{\partial t^2} =
\frac{\partial^2 \xi_{\mbox{\tiny{(2)}}}^y}{\partial t^2} = 0 \, ,
\label{gde7}
\ee
but
\be
\frac{1}{c^2} \frac{\partial^2 \xi_{\mbox{\tiny{(2)}}}^z}{\partial t^2} = \left(\partial_0 \partial_z h_{\my z0} - \onehalf \partial_z \partial_z h_{\my 00} \right) \xi_{\mbox{\tiny{(0)}}}^z.
\label{gde8}
\ee
This means that a gravitational wave does not produce movement orthogonal to the direction of propagation. In other words, it is not an orthogonal, but a longitudinal wave. Notice that, in the second order, the two degrees of freedom are represented by $h_{\my z0} = h_{\my 0z}$ and $h_{\my 00}$.
 
Considering that a detector on Earth will always be at large distances from the wave source, we use for $h_{\my \mu\nu}$ the dominant solution (\ref{dom}). Furthermore, taking into account the arbitrariness of the wave vector $K_\rho$, we can choose it in such a way that $K_0 = K_1 = K_2 = 0$. In this case, the geodesic deviation equation (\ref{gde8}) reduces to
\be
\frac{\partial^2 \xi_{\mbox{\tiny{(2)}}}^z}{\partial t^2} =
\xi_{\mbox{\tiny{(0)}}}^z \frac{\Phi_\my z \, \omega^3}{4 c} \,
\sin[2\omega(t - z/c)].
\ee
Although the wave amplitude decreases with distance, it can be assumed to be constant in the region of the experience. Accordingly, we write
\be
\frac{\partial^2 \xi_{\mbox{\tiny{(2)}}}^z}{\partial t^2} = \frac{1}{4} \,
\xi_{\mbox{\tiny{(0)}}}^z \Gamma_\my \, \omega^2
\sin[2(\omega t - z/{\lambdabar})],
\label{gde9}
\ee
where
\be
\Gamma_\my = \Phi_\my \frac{z}{\lambdabar}
\ee
represents the wave amplitude at the region of the experience, with $\lambdabar = c/\omega$ the reduced wavelength.

Observe that now the origin of the coordinate $z$ is completely arbitrary. We can then choose one of the particles to be at $z=0$, in which case $z$ will represent the position of the second particle. Assuming that the particles are initially ($t = 0$) at rest, the solution is found to be
\be
\xi_{\mbox{\tiny{(2)}}}^z = -
\frac{\xi_{\mbox{\tiny{(0)}}}^z \Gamma_\my}{16} \Big[ \sin[2(\omega t - z/{\lambdabar})] -
2 \omega t \cos[2 z/{\lambdabar}] +
\sin[2 z/{\lambdabar}] \Big].
\label{sol4}
\ee
For gravitational waves with wavelength much larger than the particle separation ($\lambdabar \gg z$), the solution becomes
\be
\xi_{\mbox{\tiny{(2)}}}^z = - \,
\frac{\xi_{\mbox{\tiny{(0)}}}^z \Gamma_\my}{16} \Big[ \sin(2 \omega t) -
2 \omega t \Big],
\label{sol5}
\ee
When a gravitational wave reaches two particles separated by a distance $\xi_{\mbox{\tiny{(0)}}}^z$ in the direction of the propagation, the distance between them will oscillate with frequency $2 \omega$, and will grow linearly with time with a velocity
\be
v = \frac{\xi_{\mbox{\tiny{(0)}}}^z \Gamma_\my \, \omega}{8}.
\ee
This behavior is the result of tidal forces produced by the passage of a gravitational wave. 

\section{Final Remarks}

Whenever use is made of a perturbation scheme, one forcibly ends up with a linear wave-equation. There is a widespread belief that gravitational waves can be approximately described by the solution of this linear wave equation. This assumption, however, is not justified. To understand it, let us make a comparison with gauge fields. As is well known, the gauge field of Chromodynamics must be nonlinear to transport color charge. Conversely, since electromagnetic waves are essentially linear, they are unable to transport their own source, that is, electric charge. Observe that, even though electromagnetic waves are linear, they do transport energy and momentum. This is possible because neither energy nor momentum is source of the electromagnetic field. As such, the energy-momentum current does not enter the electromagnetic field equation, and consequently its linearity does not restrict the energy-momentum current to be linear. Differently from electromagnetic waves, however, in order to transport energy and momentum (the source of gravitation), a gravitational wave must necessarily be nonlinear.

If we accept that the first-order equations are fully correct up to that order, and not just nearly correct as is usually assumed in the mixed approach, we arrive at the inexorable result that linear gravitational waves transport neither energy nor momentum. As a consequence, they are unable to produce any effect on free particles. One may wonder why the first-order gravitational wave, which has a non-vanishing curvature tensor, produces no effects on free particles. To get some insight on this question, let us consider the following points. First, recall that the energy-momentum current is at least quadratic in the field variables.\footnote{The classical energy of any wave, for example, is proportional to the squared amplitude.} Since it appears explicitly in the gravitational field equations, in the linear approximation it is restrict to be linear, and consequently to vanish. This means that the energy-momentum density of any linear spacetime configuration must vanish. A non-vanishing energy density can only appear in orders higher than one. This does not mean that the first-order gravitational field is physically meaningless. In fact, at the second order it will appear multiplied by itself, giving rise to nonlinear field configurations with non-vanishing energy-momentum density. Second, observe that the components of the Riemann tensor are not physically meaningful in the sense that they are different in different coordinate systems. For example, starting with the ``electric components'' $R_{i0j0}$ of the Riemann tensor, through a general coordinate transformation one can get non-vanishing ``magnetic components'' $R_{i0jk}$. By inspecting only the components, therefore, it is not possible to know whether they represent a true gravitomagnetic field, or just effects of coordinates. In order to get their physical meaning, one needs to inspect the invariants constructed out of the Riemann tensor components.\footnote{For a discussion of this point, see Ref.~\cite{CW}, page 355.} Now, as a simple calculation shows, {\em all invariants constructed out of the first-order Riemann tensor of the linear gravitational waves vanish identically}~\cite{I}. This includes the scalar curvature, the Kretschmann invariant, and the pseudo-scalar invariant. Considering that these invariants are proportional to the mass and angular momentum of the field configuration, we can conclude that the transverse components of the first-order wave are empty of physical meaning as no mass nor angular momentum (or helicity in the massless case) can be associated with them.

Motivated by the above results, we have then obtained the second-order solution to the gravitational field equations, which might represent a physical gravitational wave. Its amplitude depends explicitly on the frequency of the wave---a property typically related to nonlinearity. In contrast to the linear wave, the second-order wave is able to transport energy and momentum. This fact becomes evident if we notice that, at second order, the source energy-momentum tensor is conserved only in the covariant sense; namely, it is not really conserved~\cite{kopo}. This means that, differently from what happens at first order, at second order a mechanical system can lose energy in the form of gravitational waves. Furthermore, although the first-order field is transverse and traceless, the second order is longitudinal. This property can be understood by remembering that gravitational waves are generated, and act on particles, through tidal effects, which arise from inhomogeneities in the gravitational field. The effects they produce on free particles are then found to consist of nonlinear oscillations along the direction of propagation. This is the signature a gravitational wave will leave in a detector, the effect to be looked for.

A crucial point of these waves refer to their frequency. As is well known, the quadrupole radiation emitted by a source propagates with twice the frequency of the source (this has to do with the tidal nature of the gravitational wave generation). However, the linear gravitational wave $h_\mt^{\mu \nu}$ emerges from the perturbation scheme propagating with the same frequency of the source. To circumvent this problem, one has to artificially adjust by hands the wave frequency (see, for example, Ref.~\cite{maggiore}, page 105). On the other hand, owing to its nonlinear nature, the second order gravitational wave $h_\my^{\mu \nu}$ naturally e\-merges propagating with a frequency which is twice the source frequency, with the factor ``2'' coming from the fact that $h_\my^{\mu \nu}$ depends quadratically on $h_\mt^{\mu \nu}$. This is in agreement with the quadrupole radiation property, as well as with the tidal nature of the gravitational radiation, and is a clear indication that it is not the first-order, but the second-order wave that represents the quadrupole (physical) gravitational wave. It is also important to observe that, due to an explicit additional linear dependence on the source distance $r$, the amplitude of the dominant part of $h_\my^{\mu \nu}$ is found to fall off as $1/r$. Contrary to the usual belief, therefore, which presupposes that second-order effects fall off as $1/r^2$, second-order effects are not necessarily neglectful at large distances from the source.

It is important to remark finally that, according to Birkhoff's theorem,\footnote{For a textbook reference, see~\cite{birk}} any spherical source produces a time-independent gravitational field outside it. As a consequence, no spherically symmetric longitudinal gravitational waves can exist. However, due to the explicit dependence of the amplitude coefficient~(\ref{B2}) on the wave number, we see that the nonlinear gravitational wave considered here will never be spherically symmetric. The usual restrictions imposed by Birkhoff's theorem on longitudinal gravitational waves, therefore, do not apply to the present case of longitudinal gravitational waves. 

\section*{Acknowledgments}
The authors would like to thank FAPESP, CNPq and CAPES for partial financial 
support.


\begin{thebibliography}{99}

\bibitem{HT}
R. A. Hulse and J. H. Taylor, {\it Ap. J.} {\bf 195}, L51 (1975).

\bibitem{maggiore}
M. Maggiore, {\it Gravitational Waves} (Oxford University Press, Oxford, 2008).

\bibitem{bina}
Some reviews are: S. Bonazzola and J. A. Marck, {\it Ann. Rev. Nucl. Part. Sci.} 
{\bf 44}, 655 (1994); C. Cutler and K. S. Thorne, {\it An Overview of 
Gravitational--Wave Source}, Proceedings of the 16th International Conference on 
General Relativity and Gravitation, Durban, South Africa, 2001 [gr-qc/0204090]; 
E. E. Flanagan and S. A. Hughes, {\it New J. Phys.} {\bf 7}, 204 (2005) [gr-qc/0501041].

\bibitem{giants}
See, for example, B. Schutz, {\it Gravity from the ground up} (Cambridge University Press, Cambridge, 2003), page 313.

\bibitem{cs1}
F. I. Cooperstock, {\it Mod. Phys. Lett. A} {\bf 14}, 1531 (1999) [gr-qc/9909095];
F. I. Cooperstock, {\it Ann. Phys. (NY)} {\bf 282}, 115 (2000) [gr-qc/9904046].

\bibitem{weinberg}
See, for example, S. Weinberg, {\it Gravitation and Cosmology} (Wiley, New York, 1972).

\bibitem{mtw}
C. W. Misner, K. S. Thorne, and J. A. Wheeler, {\it Gravitation} (Freeman, New 
York, 1973).

\bibitem{salerno}
F. Canfora, G. Vilasi and P. Vitale, {\it Phys. Lett. B}  {\bf 545}, 373 (2002) [gr-qc/0205047].

\bibitem{itzu}
C. Itzykson and J. B. Zuber, {\it Quantum Field Theory} (McGraw-Hill, New York, 1980).

\bibitem{waves}
R. Aldrovandi, J. G. Pereira and K. H. Vu, {\it Found. Phys.} {\bf 37}, 1503 (2007)
[gr-qc/0709.1603].

\bibitem{living}
L. Blanchet, {\it Liv. Rev. Rel.} {\bf 5}, 3 (2002).

\bibitem{schutz}
See, for example, B. F. Schutz, {\it A First Course in General Relativity} 
(Cambridge University Press, Cambridge, 1985).

\bibitem{CW}
I. Ciufolini and J. A. Wheeler, {\it Gravitation and Inertia} (Princeton University Press, Princeton, 1995). 

\bibitem{I}
See, for example, J. Ehlers and W. Kundt, in {\it Gravitation: an introduction
to current research}, ed. by E. Witten (Wiley, New York, 1962);
S. Hawking and G. Ellis, {\it The large scale structure of space-time}
(Cambridge University Press, Cambridge, 1973), page 260;
H. Stephani, {\it General Relativity} (Cambridge University Press, Cambridge, 1982), page 164.

\bibitem{kopo}
N. P. Konopleva and V. N. Popov, {\it Gauge Fields} (Harwood, Chur, 1981).

\bibitem{birk}
R. D'Inverno, {\it Introducing Einstein's Relativity} (Clarendon Press, Oxford, 1992).

\end{thebibliography}
\end{document}